\documentclass[5p,twocolumn]{elsarticle}

\usepackage{textcomp}

\usepackage{natbib}

\usepackage{lineno,hyperref}
\modulolinenumbers[5]

\journal{Physics Letters B}

\biboptions{sort&compress}

\begin{document}

\begin{frontmatter}

\title{Measurement of the $^1S_0$ neutron-neutron effective range in
  neutron-deuteron breakup \tnoteref{preprint}}
\tnotetext[preprint]{This article is registered under preprint number: /nucl-ex/2203.02619}


\author[1,2]{R.~C.~Malone\corref{cor1}\fnref{fn1}} 
\ead{malone18@llnl.gov}

\author[1,2]{A.~S.~Crowell}

\author[1,2]{L.~C.~Cumberbatch}

\author[1,2]{B.~A.~Fallin \fnref{fn2}}

\author[1,2]{F.~Q.~L.~Friesen}

\author[1,2]{C.~R.~Howell}

\author[1,2]{C.~R.~Malone}

\author[1,2]{D.~R.~Ticehurst}

\author[1,2]{W.~Tornow}

\author[1,3]{D.~M.~Markoff}

\author[1,3]{B.~J.~Crowe}

\author[4]{H.~Wita{\l}a}

\address[1]{Triangle Universities Nuclear Laboratory, Durham, NC }

\address[2]{Department of Physics, Duke University, Durham, NC }

\address[3]{Department of Mathematics and Physics, North Carolina
  Central University, Durham, NC }

\address[4]{M. Smoluchowski Institute of
  Physics, Jagiellonian University, Krak\`{o}w, Poland}

\cortext[cor1]{Corresponding author}
\fntext[fn1]{Present address: Lawrence Livermore National
  Laboratory, Livermore, CA}

\fntext[fn2]{Present address: Savannah River National Laboratory, Aiken, SC}


\begin{abstract}
We report the most precise determination of the $^{1}S_{0}$
neutron-neutron effective range parameter ($r_{nn}$) from
neutron-neutron quasifree scattering in neutron-deuteron breakup. The
experiment setup utilized a collimated beam of 15.5~MeV neutrons and
an array of eight neutron detectors positioned at angles sensitive to
several quasifree scattering kinematic configurations. The two
neutrons emitted from the breakup reaction were detected in
coincidence and time-of-flight techniques were used to determine their
energies. The beam-target luminosity was measured in-situ with the
yields from neutron-deuteron elastic scattering.
Rigorous Faddeev-type calculations using the CD Bonn nucleon-nucleon
potential were fit to our cross-section data to determine the value of
$r_{nn}$. The analysis was repeated using a semilocal momentum-space
regularized N$^4$LO$^+$ chiral interaction potential. We obtained
values of \nobreak{$r_{nn} = 2.86 \pm 0.01 \,(stat) \pm 0.10 \,(sys)$
  fm} and \nobreak{$r_{nn} = 2.87 \pm 0.01 \,(stat) \pm 0.10 \,(sys)$
  fm} using the CD Bonn and N$^4$LO$^+$ potentials, respectively. Our
results are consistent with charge symmetry and previously reported
values of $r_{nn}$.

\end{abstract}


\begin{keyword}
  few-nucleon \sep neutron-neutron \sep effective range \sep
  neutron-deuteron breakup \sep
  neutron time-of-flight
  \href{https://doi.org/10.48550/arXiv.2203.02619}{arXiv:2203.02619}

\end{keyword}

\end{frontmatter}


Since the discovery of the neutron \cite{Cha32A,Cha32B}, much effort
has been devoted to characterizing the properties of nuclei based on
the interactions of their constituent nucleons. Due to the technical
challenges associated with quantum chromodynamics, interactions
between individual nucleons are described by effective theories
\cite{Mac11}. Modern nucleon-nucleon ($NN$) phenomenological potential
models \cite{Sto94, Wir95}, one-boson exchange models \cite{Mac96,
  Mac01}, and chiral effective theory \cite{Epe09, Mac11, Rei18} are
used to describe $NN$ scattering data. Because no direct
neutron-neutron ($nn$) scattering data exist, the isovector component
of potential models are fit to proton-proton ($pp$) scattering
data. Neutron-neutron potentials are assumed to be the same as those
for nuclear $pp$ interactions due to charge symmetry \cite{Hei32} with
small adjustments for charge-symmetry breaking effects such as the
different masses of the neutron and proton \cite{Sto94,Sto93, Mac96}.

The $^1S_0$ $nn$ scattering length ($a_{nn}$) and effective range
($r_{nn}$) have long been used to quantify charge-symmetry breaking in
the $NN$ interaction \cite{Sla89, Mil90, Mac01A}. While there have
been several recent measurements of $a_{nn}$ to resolve a
long-standing discrepancy \cite{How98, Che08, Tro99, Tro06, Huh00L,
  Huh00C}, measurements of $r_{nn}$ have not been published for over
40 years.  Early experiments that determined $r_{nn}$ from the $nn$
final state interaction (FSI) in various scattering systems resulted
in a large spread of values (2.0 - 3.2~fm) with large uncertainties
(20 - 60\%) \cite{Bau66, Gro70, Lar70, Slo71, Zei72, Zei74,
  Kec75}. Gabioud \textit{et al.}  extracted both $a_{nn}$ and
$r_{nn}$ based on measurements of the photon energy spectrum from the
reaction $^2H(\pi^-,\gamma)2n$ and achieved the most precise result to
date, \nobreak{$r_{nn} = 2.80 \pm 0.11 \, (exp) \pm 0.11 \,
  (theory)$}, consistent with charge symmetry
\cite{Gab79,Gab81,Gab84}. More recently, $r_{nn}$ was calculated from
the value of $a_{nn}$ determined from measurements of the $nn$ FSI in
neutron-deuteron ($nd$) breakup \cite{Bab12, Zuy09, Kon10,
  Kon11}. These determinations are not ideal because the low relative
momentum between the neutrons in the $nn$ FSI makes this configuration
much more sensitive to $a_{nn}$ than to $r_{nn}$.

Neutron-neutron quasifree scattering (QFS) in the $nd$ system is ideal
for measuring $r_{nn}$. In this kinematic configuration, the momentum
of the incident neutron is transferred exclusively to the neutron in
the deuteron, i.e., the proton remains at rest in the laboratory frame
during the scattering process. The cross section for $nn$ QFS in $nd$
breakup is highly sensitive to $r_{nn}$ and is insensitive to
off-shell effects of the $NN$ potential, three-nucleon forces, and
$a_{nn}$ \cite{Wit10, Wit11}. For this reason, several early
experiments determined $r_{nn}$ from measurements of $nn$ QFS in $nd$
breakup. These experiments were performed at incident neutron energies
between 14 and 25 MeV \cite{Sla71, Bov78, Sou79, Gur80,
  Wit80}. Although these measurements agreed well with theory, they
were limited by large statistical uncertainties and were compared to
theory which implemented several simplifying assumptions and used
phenomenological $NN$ interactions. A weighted average of the results
gives a value of \nobreak{$r_{nn} = 2.68 \pm 0.16$ fm}, consistent
with the value from Gabioud \cite{Sla89}. The recommended value of
\nobreak{$r_{nn} = 2.75 \pm 0.11 $ fm} is a weighted average of all
previous measurements, but the value is dominated by the $\pi^{-}d$
and $nn$ QFS experiments \cite{Sla89}. The average value is consistent
with the charge-symmetric proton-proton value of \nobreak{$r_{pp} =
  2.85 \pm 0.04$ fm} \cite{Sla89, Mil90, Mac01A}.

The situation was complicated by three recent measurements of $nn$ QFS
in $nd$ breakup at incident neutron energies of 10.3, 26, and 25 MeV
\cite{Lub92,Sie02,Rua07}. Rigorous Faddeev calculations \cite{Glo96}
using the CD Bonn potential \cite{Mac96} underpredicted the measured
cross section by about 16\% \cite{Sie02, Rua07, Wit11}; the theory
predicted the shape of the data but not its magnitude.  Inclusion of
three-nucleon ($3N$) forces did not significantly change the predicted
cross section \cite{Wit10}. Experimentally, a large error ($\approx
16$\%) in the determination of the beam-target luminosity would
explain the discrepancy. On the theory side, the discrepancy could be
removed by scaling the $^{1}S_{0}$ $nn$ interaction matrix element by
a factor of 1.08. However, this remedy drastically alters
the value of $r_{nn}$, resulting in a significant charge-symmetry
breaking effect \cite{Wit11}.

We have performed new measurements of the cross section for $nn$ QFS
in $nd$ breakup to investigate the discrepancy reported in
Refs. \cite{Lub92,Sie02,Rua07}. In contrast to previous experiments,
the integrated beam-target luminosity was determined from the yields
for $nd$ elastic scattering. This removes several sources of
systematic uncertainty because the $nd$ elastic scattering yields are
measured simultaneously with the breakup yields and the absolute
neutron flux and number of deuterium nuclei in the target do not need
to be known independently. This technique has been successfully
implemented in a previous measurement \cite{Mal20}. The present
experiment was conducted at a different energy with more
configurations sensitive to $nn$ QFS than previous work
\cite{Sie02,Rua07}. In this letter, we discuss the setup of the
experiment, the data-analysis procedures, and the results for the
kinematic configurations sensitive to $r_{nn}$. More details about the
experiment and results from other $nd$ breakup configurations will be
discussed in a forthcoming publication.


\begin{figure}[h]
\centering
\includegraphics[width=\linewidth]{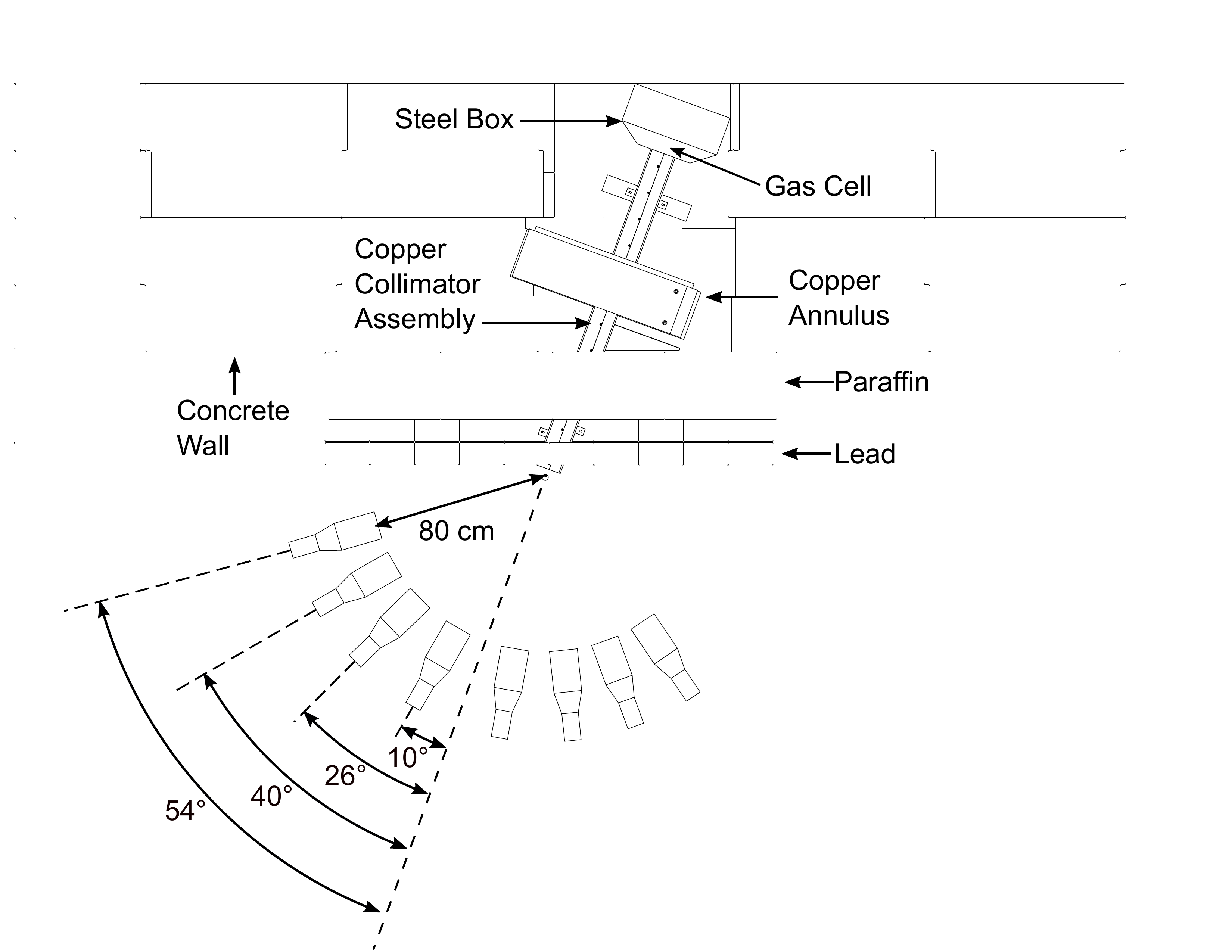}
\caption{\label{fig:ssa-setup} A diagram of the experiment setup
  (distances are to scale). The CD$_2$ sample is 177.6~cm from the
  center of the neutron production cell, and the detectors are located
  80~cm from the target (center-to-center distance) at nominal angles
  of 10\textdegree{}, 26\textdegree{}, 40\textdegree{}, and
  54\textdegree{} on either side of the beam axis. More details are
  given in the text.}
\end{figure}

Measurements were performed at the tandem accelerator facility at the
Triangle Universities Nuclear Laboratory. The experiment setup is
shown in Fig. \ref{fig:ssa-setup}. A beam of deuterons was directed
into a 7.26-cm long cell of deuterium gas pressurized to 7.1 atm to
produce neutrons at $15.5 \pm 0.25$ MeV (full width) via the
$^{2}$H($d,n$)$^{3}$He reaction. The incident deuteron beam was pulsed
(T~=~400~ns, $\Delta$t~=~2~ns~FWHM) and the arrival of each beam pulse
was detected by a capacitive beam pickoff unit immediately upstream of
the deuterium gas cell. This provided a time reference for neutron
time-of-flight (TOF) measurements. The average deuteron beam current
on target was kept at 800~nA.

The use of a copper collimator surrounded by a large shielding wall
resulted in a rectangular neutron beam with a plateau of constant flux
36~mm wide $\times$ 55~mm high at the location of the CD$_2$ target,
which was suspended 177.6~cm downstream from the center of the neutron
production cell.

Eight BC-501A liquid organic scintillators were used for neutron
detection. The detectors were right cylinders (diameter = 12.7~cm,
thickness = 5.08~cm) oriented with symmetry axes pointing at the
scattering sample. Detectors were placed approximately 80~cm from the
scattering sample (center-to-center distance) at nominal angles of
10\textdegree{}, 26\textdegree{}, 40\textdegree{}, and 54\textdegree{}
on opposite sides of the neutron beam axis. The exact positions and
scattering angles for each detector were determined from a survey of
the detectors and scans of the neutron beam profile. Another
cylindrical BC-501A scintillator (diameter = 3.81~cm, thickness =
3.81~cm) was placed in the neutron beam 507.4~cm downstream from the
neutron production cell to monitor the neutron beam flux during data
collection (not shown in Fig.  \ref{fig:ssa-setup}).

To reduce background events, two cuts were applied to the data. A
pulse-height threshold equal to one-half the energy of the $^{137}$Cs
Compton-scattering edge (239 keV-electron-equivalent) was used to
reduce backgrounds from low-energy particles. Pulse-shape
discrimination (PSD) techniques were used to reduce backgrounds from
gamma-ray events.

Two right cylinders composed of deuterated polyethylene (CD$_2$,
Cambridge Isotope Laboratories, Inc., DLM-220-0) and graphite were
used as scattering samples. Physical properties of the samples are
given in Table \ref{tab:sample-dimensions}. Each sample was mounted in
the beam with their symmetry axes vertical. The entire volume of each
sample was within the area of constant neutron flux. The graphite
sample was used to measure backgrounds from neutron scattering on
carbon in the CD$_2$ sample. Other backgrounds such as neutron
scattering from air were measured with an empty sample holder.

\begin{table}[h]
    \caption{\label{tab:sample-dimensions}Properties of the scattering
      samples used.}
\centering
\begin{tabular}{lccc}
  \hline
  Sample & Mass (g) & Diameter (mm) & Height (mm)\\
  \hline
  CD$_{2}$          &  25.172  &  28.3  & 36.4 \\
  Graphite   &  42.055  &  28.6  & 38.0 \\
  \hline
\end{tabular}
\end{table}

To avoid unacceptably high dead times of the data-acquisition system
(DAQ), a trigger circuit with separate branches for single detector
events and coincidence events was used. Coincidences between detectors
on opposite sides of the neutron beam ($\Delta \phi=180^{\circ}$)
within a window of 850~ns triggered the DAQ. The single-event branch
trigger rate was divided by a factor of 10, delayed 400~ns, and vetoed
by the coincidence branch before triggering the DAQ. This allowed all
coincidence events to be measured with a reasonable DAQ dead time
($\sim 10\%$) while accumulating sufficient counts from \textit{nd}
elastic scattering to achieve a statistical accuracy better than 0.1\%
in the beam-target luminosity determination.

\begin{figure}[!ht]
\centering
\includegraphics[width=\linewidth]{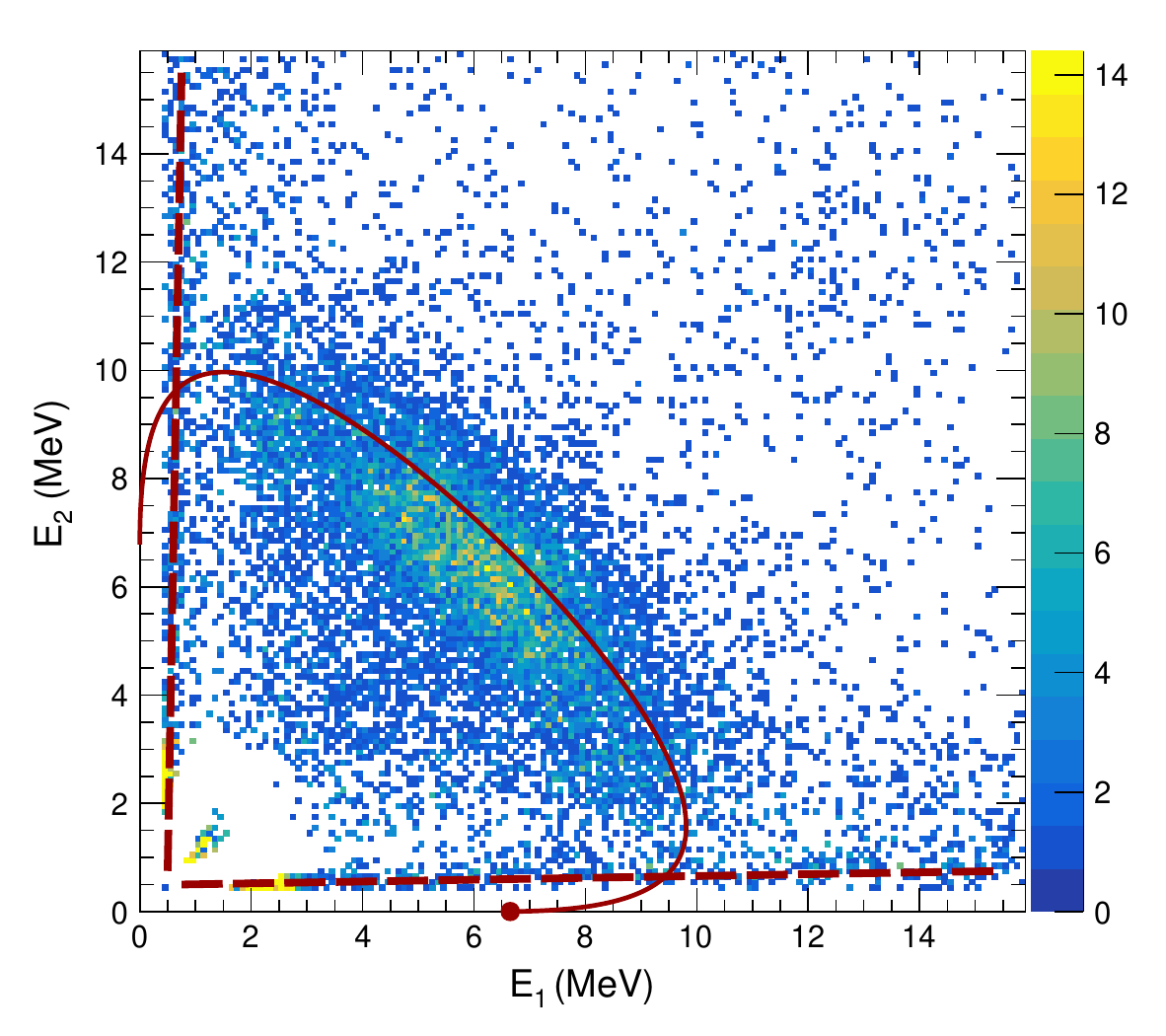}
\caption{\label{fig:energy-2d-raw} Two-dimensional coincidence energy
  spectrum for the detectors at 40\textdegree{} on opposite sides of
  the beam axis. The accidental spectrum has been subtracted from the
  raw spectrum. The red curve is the ideal kinematic locus, and the red
  dot marks the point where $S=0$. The red dashed lines indicate
  regions affected by cross-talk events. The counts in the bins at low
  energy ($E_1$ and $E_2 < 3$ MeV) exceed the vertical scale and may
  contain 15 - 60 counts.  }
\end{figure}

Two types of coincidence spectra were measured: (1) the raw
coincidence spectrum containing true and accidental coincidences, and
(2) the accidental coincidence spectrum. Coincidences between events
originating from two consecutive beam pulses ($\Delta t \sim 400$~ns)
were used to measure the accidental coincidence spectra. The
coincidence spectrum for the pair of detectors at 40\textdegree{} on
opposite sides of the beam is shown in
Fig. \ref{fig:energy-2d-raw}, where accidental coincidence spectrum has
been subtracted from the raw coincidence spectrum for display.

A background due to detector cross talk, in which a single neutron
scattered between two detectors and was detected in both, contributed
to the raw coincidence spectrum. These events were separated by less
than 100~ns in time and were indistinguishable from real $nn$
coincidences. However, due to the experiment geometry, these events
fell outside the region of the data reported here. Coincidences due to
cross talk form the bands indicated by the red dashed lines in
Fig. \ref{fig:energy-2d-raw}. The events around $E_{1} = E_{2} =
1.2$~MeV are due to cross-talk coincidences from
gamma rays not removed by the PSD cut.

Coincidence events from $nd$ breakup fall on a locus of kinematically
allowed neutron energies. The red curve in
Fig. \ref{fig:energy-2d-raw} is the ideal locus, or $S$ curve, defined
by the central geometry of the detector pair. The variable $S$
measures the arc length along the curve in a counterclockwise
direction beginning at the point where the energy of the second
neutron reaches a minimum \cite{Glo96}, indicated by the red dot
in Fig. \ref{fig:energy-2d-raw}. For each detector pair, the raw and
accidental coincidence events in a band around the $S$ curve were
projected onto the ideal kinematic locus. The $S$ curve was divided
into 0.5~MeV-wide bins and each detected event was projected into the
nearest bin on the $S$ curve. The projected accidental coincidence
spectrum was subtracted from the projected raw spectrum for each
detector pair to obtain the true $nn$ coincidence yields as a function
of $S$. The breakup cross section was computed using these yields.


The present data were compared to rigorous ab-initio three-body
calculations using the CD-Bonn potential \cite{Mac96, Mac01} and the
semilocal momentum-space (SMS) regularized N$^4$LO$^+$ chiral
interaction of the Bochum group \cite{Rei18} with the cutoff
$\Lambda=450$~MeV in the Faddeev formalism applying the technique
described in Ref. \cite{Glo96}. A Monte-Carlo (MC) simulation of the
experiment was used to average the point-geometry theory predictions
over the energy spread and finite geometry of the experiment to allow
accurate comparison between theory and data. The MC simulation was
also used to determine the average values of the product of neutron
detector efficiencies and neutron transmission probabilities as a
function of $S$, which were necessary to compute the breakup cross
section. Finally, the simulation was used to quantify contributions of
background processes to neutron scattering yields. The details of the
simulation can be found in Ref. \cite{Mal20}.

Background processes that were simulated for both breakup and $nd$
elastic scattering included: multiple scattering of neutrons in the
sample, in-scattering of neutrons from shielding materials and
adjacent detectors, and neutrons produced via $^2H(d,n)^3He$ on
deuterons implanted in the gold beam stop at the end of the gas
cell. Two additional backgrounds for $nd$ elastic scattering were
simulated: neutron scattering from the 1.6\% hydrogen impurity in the
CD$_2$ sample, and $nd$ breakup events in which only one neutron was
detected. The fraction of background events determined with the MC
simulation was subtracted from the measured $nn$ coincidence yields
and the $nd$ elastic scattering yields. The corrections for elastic
scattering were between 7-12\% and the average correction for breakup
events varied from 6-8\%, depending on the scattering configuration.


The yields from \textit{nd} elastic scattering in each detector were
used to determine the integrated beam-target luminosity. Backgrounds
from neutron scattering on carbon and air were measured using the
graphite sample and an empty target holder, respectively.
The TOF spectrum for each detector accumulated with the empty target
holder was normalized and subtracted from the spectra measured with
the CD$_2$ and graphite samples. The empty-target TOF spectra were
normalized using the integrated beam current (BCI), the gas pressure
in the neutron production cell, and the DAQ live time fraction. The
TOF spectra measured with the graphite sample were normalized to the
spectra measured with the CD$_2$ sample in a similar way. The
normalization factor also included the ratio of carbon nuclei in the
two samples determined using data from a previous experiment
\cite{Mal20}. The normalized graphite TOF spectra were subtracted from
the CD$_2$ TOF spectra to obtain the raw yields from $nd$ elastic
scattering. Also, the raw yields were corrected for backgrounds
quantified with the MC simulation.

The luminosity per BCI measured by all eight detectors agreed with a
standard deviation of 2.1\%. The geometric mean of the beam-target
luminosity measured by all detectors except those at 10\textdegree{}
was used to compute the breakup cross section. The $nd$ yields
measured by the detectors at 10\textdegree{} were excluded from the
luminosity determination because of the large uncertainty in
subtracting the background contributions ($\sim 70\%$) due to neutron
scattering on carbon and air.


Several $nn$ QFS configurations were measured using detectors
positioned on opposite sides of the beam axis. The detector pair at
$\theta_1 = \theta_2 = 40$\textdegree{} measured an exact
quasifree-scattering configuration. The detector pair at $\theta_1 =
\theta_2 = 26$\textdegree{} and the two pairs at $\theta_1 =
26$\textdegree{}, $\theta_2 = 40$\textdegree{}, measured
configurations near $nn$ QFS, where the proton energy reaches a
minimum of 0.4 MeV and 0.1 MeV, respectively. Two other $nn$ QFS
configurations were measured; however, the cross sections for those
configurations are dominated by $np$ final-state interactions and
therefore were not used to determine $r_{nn}$.

\begin{figure}[!h]
  \centering
\includegraphics[width=0.95\linewidth]{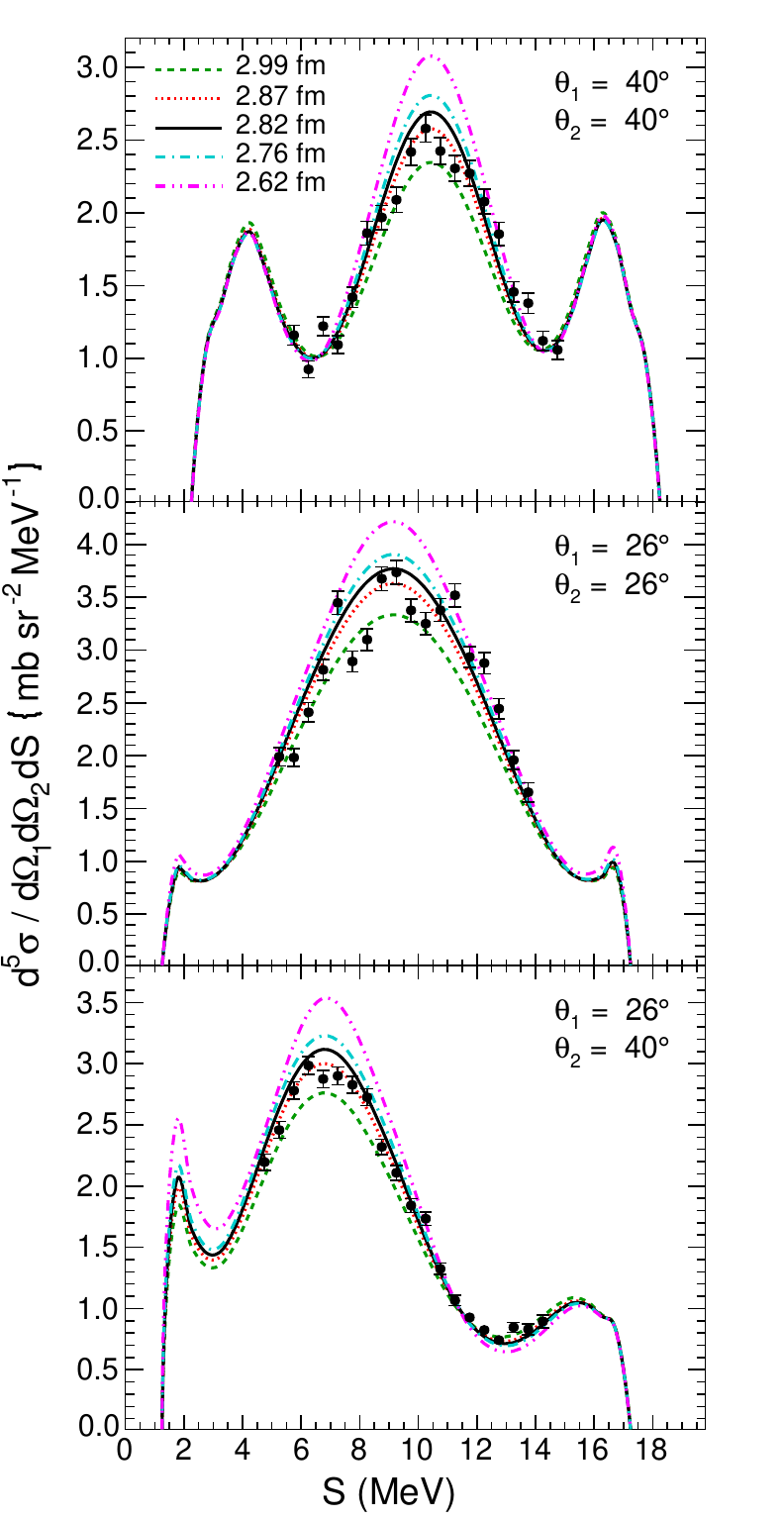}
\caption{\label{fig:nnqfs-xsecs} Plot of the measured $nd$ breakup
  cross section as a function of $S$ for the $nn$ QFS configurations
  most sensitive to $r_{nn}$. The error bars represent statistical
  uncertainties only; there is a systematic uncertainty of $\pm
  6.3\%$. The solid curve is the result of the MC simulation using the
  standard CD Bonn potential. Other curves represent the MC simulation
  result using the CD Bonn potential with values of $r_{nn}$ indicated
  by the legend in the top panel. }
\end{figure}

Our results for the $nn$ QFS cross section are presented in
Fig. \ref{fig:nnqfs-xsecs}. The measured data are given by the points
and the error bars represent statistical uncertainty. Not shown on the
plot is a systematic uncertainty of $\pm 6.3\%$. The data points for
the configuration at $\theta_1 = 26$\textdegree{}, $\theta_2=
40$\textdegree{} are a statistically weighted average of the results
from the two detector pairs. The curves in Fig.
\ref{fig:nnqfs-xsecs} represent the result of the MC simulation using
$nd$ breakup cross sections calculated with the CD Bonn potential with
different values of $r_{nn}$. The curves provide a representative
sample covering the full range of the values $r_{nn}$ used in the
analysis.

\clearpage

\begin{table}[h]
    \caption{\label{tab:nnqfs-sys-unc} Sources of systematic
      uncertainty in the determination of the cross section for $nn$
      QFS in $nd$ breakup. All uncertainties are given as one standard
      deviation.} \centering
    \begin{tabular}{lr}
      \hline
      Source & Magnitude (\%) \\
      \hline
      Coincidence yields                       &  1.5 \\
      Elastic scattering yields                &  2.8 \\
      Efficiency normalization                 &  2.9 \\
      Efficiency shape                         &  3.0 \\
      Detector gain drift                      &  1.0 \\
      Neutron transmission                     &  2.7 \\
      $nd$ elastic cross section               &  1.6 \\
      Detector solid angle                     &  0.9 \\
      Single event live time                   &  0.1 \\
      Coincidence event live time              &  0.6 \\
      \hline
      Total                                    & 6.3 \\
      \hline
    \end{tabular}
\end{table}

Sources of systematic uncertainty are listed in Table
\ref{tab:nnqfs-sys-unc}. The uncertainty in the measured $nn$
coincidence yields is mainly due to the MC corrections for multiple
scattering and in-scattering, but also includes contributions from the
accidental background subtraction and neutrons produced at energies
below 15.5~MeV via $(d,n)$ reactions on contaminants in the gas
cell. Subtraction of the graphite spectra and simulated backgrounds
contribute the most to the uncertainty in the extracted elastic
scattering yields. The MC simulation uses as input simulated detector
efficiency curves fit to measurements of the neutron flux from the
$^{2}$H($d,n$)$^{3}$He reaction at zero degrees
\cite{Die82,Dro15}. The uncertainty in the overall normalization and
shape of the efficiency curve were determined from those measurements
as in Ref. \cite{Mal20}. The effect of small changes in the detector
gain on the simulated detector efficiency was estimated using
variations of the efficiency curve due to changes in the detector
threshold setting. Neutron transmission factors were calculated with
the MC simulation using cross sections and associated uncertainties
from ENDF/B-VII.1 \cite{Cha11}. The $nd$ elastic scattering cross
section was computed using the CD Bonn potential, and the error is
estimated as the difference between predictions of several modern $NN$
potentials \cite{Ski18}. The uncertainty in the detector solid angle
is due to the precision of measuring the positions of the detector
faces (0.1 cm). The difference between multiple methods used to
determine the DAQ live-time fraction is used as an estimate of its
uncertainty.

To determine $r_{nn}$, the MC simulation was run using breakup cross
sections calculated with the CD Bonn $nn$ $^{1}S_{0}$ matrix element
scaled by seven different factors, resulting in different values of
$r_{nn}$. Scaling this matrix element also alters $a_{nn}$; however,
because the $nn$ QFS cross section is sensitive only to variations in
the effective range and not the scattering length, this is adequate to
determine $r_{nn}$ \cite{Wit11}. For every detector configuration, the
value of $\chi^2$ was computed as a function of $r_{nn}$ using the
integral of the cross section over the QFS peak where the sensitivity
to changes of $r_{nn}$ is greatest. A second-degree polynomial was fit
to the $\chi^2$ function and the minimum of this fit was taken as the
best value of $r_{nn}$. The statistical uncertainty in the extracted
value of $r_{nn}$ is given by $ \Delta r_{nn} = | r_{nn}(\chi^2_{min}
+ 1) - r_{nn}(\chi^2_{min}) | $. The integral of the QFS theoretical
cross section as a function of $r_{nn}$ was used to convert the
systematic uncertainty in the cross section to uncertainty in
$r_{nn}$. The analysis was repeated using the SMS chiral N$^4$LO$^+$
potential of the Bochum group \cite{Rei18} with the cutoff
$\Lambda=450$~MeV. Our results are summarized in Table
\ref{tab:rnn-vals}. The values of $r_{nn}$ determined from the three
$nn$ QFS configurations agreed within statistical uncertainties. The
final result is given as a weighted average of those values. The
values extracted using the two different potentials agree well, and
our values are consistent with the recommended value $r_{nn} = 2.75
\pm 0.11$ fm and the charge-symmetric value of $r_{pp} = 2.85 \pm
0.04$ fm \cite{Sla89, Mil90, Mac01A}.

\begin{table}[!htb]
  \caption{\label{tab:rnn-vals} Values of $r_{nn}$ determined from
    different angular configurations in this experiment using the
    CD-Bonn potential. The last two rows give the weighted average
    determined using the CD Bonn and N$^4$LO$^+$ potentials. All
    uncertainties are given as one standard deviation. See text for
    details.}  \centering
  \begin{tabular}{lr}
    \hline
    Configuration  & $r_{nn} \pm \sigma_{stat} \pm \sigma_{sys}$ (fm) \\

    \hline
    $\theta_1=40$\textdegree{} $\theta_2=40$\textdegree{} &
    $2.85 \pm 0.02 \pm 0.09 $ \\
    $\theta_1=26$\textdegree{} $\theta_2=26$\textdegree{} &
    $2.85 \pm 0.02 \pm 0.11  $  \\
    $\theta_1=26$\textdegree{} $\theta_2=40$\textdegree{} &
    $2.87 \pm 0.01  \pm 0.10  $ \\
    \hline
    CD Bonn Average    &
    $2.86 \pm 0.01 \pm 0.10  $  \\
    N$^4$LO$^+$ Average & $2.87 \pm 0.01 \pm 0.10 $ \\
    \hline
  \end{tabular}
\end{table}

We have performed measurements of the cross section for $nn$ QFS in
$nd$ breakup using a new technique to determine the integrated
beam-target luminosity based on the $nd$ elastic scattering yields
measured simultaneously with the $nn$ coincidences from $nd$ breakup.
Our results, summarized in Table \ref{tab:rnn-vals}, provide the first
measurement of the $nn$ effective range parameter using modern $NN$
potentials and the most precise determination from $nn$ QFS in $nd$
breakup. The data also suggest that the previously reported
discrepancies between theory and data in the $nn$ QFS cross section at
10.3, 26 and 25 MeV \cite{Lub92, Sie02, Rua07} may be due to
systematic errors in the determination of the beam-target luminosity
leading to incorrect normalization of the breakup cross
section. Another possibility is that the discrepancy is energy
dependent and only becomes evident in measurements at incident neutron
energies above 20~MeV. Further measurements at higher incident neutron
energies should be carried out to investigate this possibility.

\section*{Acknowledgments}
  The authors thank the TUNL technical staff for their
  contributions. The authors appreciate the use of the supercomputer
  cluster of the JSC in J{\"u}lich, Germany, where parts of the
  numerical calculations were performed. This work is supported in
  part by the U.S. Department of Energy under grant
  Nos. DE-FG02-97ER41033 and DE-SC0005367 and by the Polish National
  Science Center under grant No. DEC-2016/22/M/ST2/00173. Part of this
  work was performed under the auspices of the U.S. Department of
  Energy by Lawrence Livermore National Laboratory under Contract
  DE-AC52-07NA27344.

\bibliography{RCMalone_nnqfs_15.5MeV_plb}

\end{document}